\documentclass[aps,pra,showpacs,twocolumn]{revtex4}
\usepackage{graphicx}
\usepackage{bm}
\begin{document}
\preprint{}
\title{Contradictory uncertainty relations}
\author{Alfredo Luis}
\email{alluis@fis.ucm.es}
\homepage{http://www.ucm.es/info/gioq}
\affiliation{Departamento de \'{O}ptica, Facultad de Ciencias
F\'{\i}sicas, Universidad Complutense, 28040 Madrid, Spain }
\date{\today}

\begin{abstract}
We show within a very simple framework that different measures of 
fluctuations lead to uncertainty relations resulting in contradictory 
conclusions. More specifically we focus on Tsallis and R\'{e}nyi 
entropic uncertainty relations and we get that the minimum 
uncertainty states of some uncertainty relations are the maximum 
uncertainty states of closely related uncertainty relations, and 
vice versa.
\end{abstract}


\maketitle

\textit{Introduction.--} 
Uncertainty relations is a rather basic issue in quantum physics.
This point has been mostly addressed in terms of the product of 
variances of the corresponding Hermitian operators representing 
observables. Nevertheless, there are situations where such 
formulation is not satisfactory enough and alternative approaches 
are required. For example: (i) variance is not always a well 
behaved estimator of fluctuations beyond Gaussian statistics 
\cite{var}, (ii) in finite-dimensional systems there are no 
nontrivial lower bounds for the product of variances, since 
the variance of an observable can vanish while the variance of 
any other is bounded from above \cite{ypra}, (iii) for periodic 
variables such as angle and phase variance is ambiguous and 
rather useless by strongly depending on the angle/phase window 
\cite{fase}, and (iv) there are observables not easily represented 
by Hermitian operators \cite{fase}. This has prompted the 
introduction of alternative measures of fluctuations and 
uncertainty relations \cite{ypra,fase,atv,Ts,PP,Ry,yR,ZZ}. 

The question addressed in this work is that different assessments 
of fluctuations may lead to uncertainty relations resulting in 
contradictory conclusions. This holds even within the same family 
of uncertainty measures, such as Tsallis and R\'{e}nyi entropies
\cite{Ts,PP,Ry}. These contradictions are quite relevant given 
the importance of quantum uncertainty relations, from fundamental 
issues to metrological applications. 

\textit{Tsallis and R\'{e}nyi entropies.--} 
For definiteness we will consider the Tsallis entropies \cite{Ts}
\begin{equation}
S_q (A) = \frac{1-\sum_j p_j^q}{q-1}   ,
\end{equation}
where $p_j$ is the probability of the outcome $j$ of the observable 
$A$, and $q$ is a real parameter. Note that $S_q (A)$ is always 
nonnegative. This is a suitable measure of uncertainty. Minimum 
uncertainty $S_q (A) =0$ holds when all the probability is 
concentrated in a single outcome $p_j = \delta_{j,k}$ for any $k$, 
so that $\sum_j p_j^q =1$. Maximum uncertainty occurs when all the 
outcomes are equally probable $p_j= 1/N$ where $N$ is the number 
of outcomes.   

This family includes the Shannon entropy in the limit $q 
\rightarrow 1$
\begin{equation}
S_{q \rightarrow 1} (A) = - \sum_j p_j \ln p_j .
\end{equation}
This is also closely related to R\'{e}nyi entropy, that we will 
express as \cite{ypra,Ry,yR}
\begin{equation}
R_q (A) = \left ( \sum_j p_j^q \right )^{1/(1-q)} ,
\end{equation}
so that for Gaussian-like variables $R_q (A) \propto \Delta A$.
This takes its minimum $R_q (A) =1$ when all the probability is 
concentrated in a single outcome  $p_j = \delta_{j,k}$, while 
the maximum $R_q (A) =N$ occurs when all the outcomes are equally 
probable $p_j= 1/N$, where $N$ is the number of outcomes.   

The Tsallis entropies also include the variance $( \Delta A)^2$ of 
two-outcome observables within two-dimensional spaces, with $A$ 
represented by the Hermitian operator 
\begin{equation}
A = | a \rangle \langle a |- | \neg a \rangle \langle \neg a |,
\end{equation}
with $ \langle a |\neg a \rangle =0$, since for $q=2$ we have
\begin{equation}
S_2 (A) = 2 p_a (1-p_a) = \frac{1}{2} ( \Delta A)^2 ,
\end{equation}
with $p_a = \langle a | \rho | a \rangle$ for any state $\rho$. 

These measures may enter in uncertainty relations for two observables
$A,B$ via nontrivial lower bounds on different combinations of these 
entropies. The most frequent combinations considered in the literature 
\cite{atv,Ts,PP,Ry,yR,ZZ} are of the sum of Tsallis entropies 
\begin{equation}
\Sigma_q = S_q (A) + S_q (B) ,
\end{equation}
the product of R\'{e}nyi entropies
\begin{equation}
\Pi_q = R_q (A) R_q (B) ,
\end{equation}
and the combination of Tsallis entropies proposed in Ref. $U_q$ 
\cite{PP}
\begin{equation}
U_q = S_q (A) + S_q (B) + (1-q) S_q (A) S_q (B) .
\end{equation}
For the sake of symmetry we are going to consider the same parameter
$q$ for both $A$ and $B$.

In this work we are not interested in the precise lower bounds of 
$\Sigma_q$, $\Pi_q$, or $U_q$. Instead we are worried by 
contradictions between the conclusions derived from different 
choices of $q$ for the same family of entropy combinations. 

\textit{Two-dimensional observables.--}
To reveal contradictions as simply as possible we consider a 
two-dimensional system and two observables $A,B$ with outcomes 
$A=(a, \neg a)$, $B=(b, \neg b)$, and probabilities $p_k$, 
$k =a, \neg a, b, \neg b$, given by projection of the system 
state $| \psi \rangle$ (assumed pure for simplicity) on the 
corresponding vectors $| k \rangle$
\begin{equation}
p_k = \left | \langle k | \psi \rangle \right |^2 ,
\end{equation}
with $p_{\neg k} = 1 - p_k$ and $\langle \neg k | k \rangle =0$. 

In the general case the states $| a \rangle$ and $| b \rangle$ 
will not be orthogonal so that 
\begin{equation}
| b \rangle = \cos \delta | a \rangle + \sin \delta | \neg a \rangle .
\end{equation}
For definiteness we will consider $\pi/4 \geq \delta \geq 0$ since 
otherwise we may exchange $a \leftrightarrow \neg a$ for example. 

The case $\delta = \pi/4$ corresponds to typical complementary 
observables, so that for $| \psi \rangle = | b \rangle$ there is 
$p_{\neg a} = p_a = 1/2$  and vice versa. For example this is the case 
of two orthogonal components of an 1/2 spin, say $A= \sigma_z$ and 
$B= \sigma_x$, where $\sigma_{x,z}$ are the corresponding Pauli matrices.

\textit{Comparison of uncertainty relations for complementary 
observables.--}
For definiteness let us consider system states in the form 
\begin{equation}
| \psi \rangle = \cos \theta | a \rangle + \sin \theta | \neg a 
\rangle ,
\end{equation}
where $\theta$ is a variable, so that 
\begin{equation}
p_a = \cos^2 \theta , \qquad  p_b = \cos^2 (\theta - \delta ) .
\end{equation}

In Figs. 1,2, and 3 we plot $\Pi_q$, $U_q$, and $\Sigma_q$
as functions of $\theta$ for $\delta = \pi/4$ and several 
values of $q$. It can be appreciated that in all the cases 
there is a maximum/minimum exchange depending on the value 
of $q$. Moreover, in Fig. 4 we plot the second derivative 
of $\Pi_q$, $U_q$, and $\Sigma_q$ at $\theta= \delta /2 =
\pi/8$ 
\begin{equation}
F^{\prime \prime} = \left . \frac{d^2 F}{d \theta^2} 
\right |_{\theta = \delta/2}, \qquad F=  \Pi_q, \Sigma_q, U_q,
\end{equation}
as functions of $q$ showing the change from maximum (negative 
$F^{\prime \prime}$) to minimum (positive $F^{\prime \prime}$). 
For example for $\Sigma_q$ the exchange holds for $q$ between 
$q=2$ and $q=3$.

The states disputing the maxima and minima are 
\begin{equation}
| \psi_{\theta = \delta/2} \rangle \propto | a \rangle + | b 
\rangle ,
\end{equation}
that maximizes the product of probabilities $p_a p_b$ with $p_a 
= p_b$ and $S_q (A) = S_q (B)$ \cite{yomlh}, 
\begin{equation}
| \psi_{\theta = \delta/2 + \pi/4} \rangle \propto |\neg  a 
\rangle + | b \rangle ,
\end{equation}
that maximizes the product of probabilities $p_{\neg a} p_b$ 
with $p_{\neg a} = p_b$ and $S_q (A) = S_q (B)$, and the state 
associated to $\theta = 0,\delta$ $\mathrm{modulus} (\pi/2)$ 
\begin{equation}
| a \rangle, | \neg a \rangle, | b \rangle, | \neg b \rangle ,
\end{equation}
that corresponds to either $p_{a, \neg a}=1$ with $S_q (A)=0$ 
and $\Delta A =0$, or $p_{b, \neg b} =1$ with $S_q (B)=0$ and 
$\Delta B=0$.

\begin{figure}
\begin{center}
\includegraphics[width=6cm]{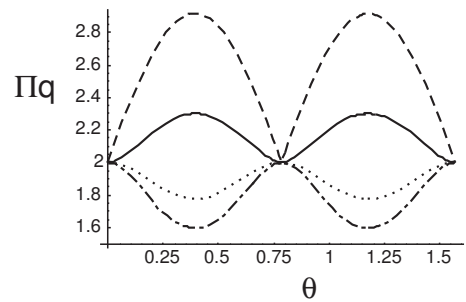}
\end{center}
\caption{Plot of $\Pi_q = R_q (A) R_q (B)$ as a function of 
$\theta$ for $\delta = \pi/4$ and $q=0.5$ (dashed line), $q=1$ 
(solid line), $q=2$ (dotted line) and $q=3$ (dash-dotted line).}
\end{figure}

\begin{figure}
\begin{center}
\includegraphics[width=6cm]{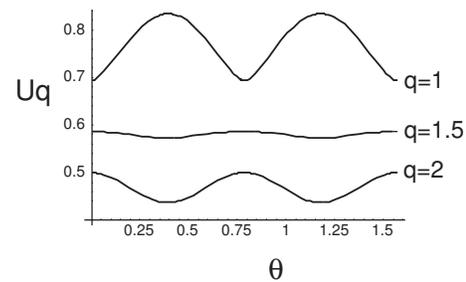}
\end{center}
\caption{Plot of $U_q$ as a function of $\theta$ for $\delta = \pi/4$ 
and $q=1,1.5,2$.}
\end{figure}

\begin{figure}
\begin{center}
\includegraphics[width=6cm]{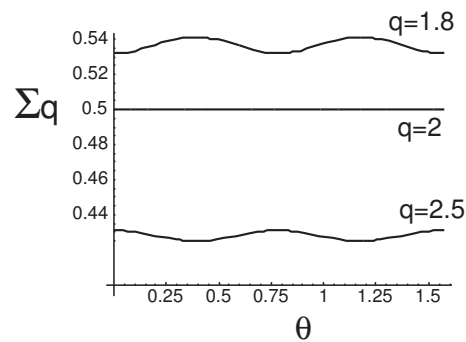}
\end{center}
\caption{Plots of $\Sigma_q = S_q (A) + S_q (B)$ as a function 
of $\theta$ for $\delta = \pi/4$ and $q=1.8, 2, 2.5$.}
\end{figure}

\begin{figure}
\begin{center}
\includegraphics[width=6cm]{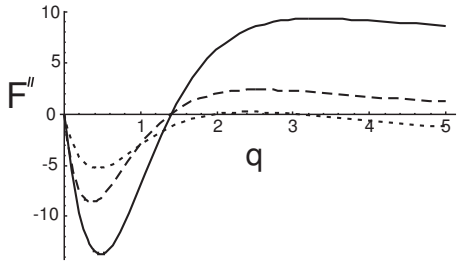}
\end{center}
\caption{Plot the second derivative at $\theta = \delta/2$ of 
$\Pi_q$ (solid line), $U_q$ (dashed line), and $\Sigma_q$ 
(dotted line) as functions of $q$ for $\delta = \pi/4$.}
\end{figure}

\textit{Comparison of uncertainty relations for not fully 
complementary observables.--} The above behavior is reproduced 
also when considering not completely complementary observables,
this is  $\delta \neq \pi/4$, appearing also some new features.
In Figs. 5, 6, and 7 we plot $\Pi_q$, $U_q$, and $\Sigma_q$ 
as functions of $\theta$ for $\delta = 0.7$ and several 
values of $q$. It can be appreciated that a local maximum at  
$\theta = \delta/2$ for lower $q$ becomes absolute minimum 
for larger $q$. Accordingly, the absolute minima at $\theta =0,
\delta$ for low $q$ are no longer minima for larger $q$. In 
Fig. 8  we plot the second derivative $\Pi_q$, $U_q$, and 
$\Sigma_q$ at $\theta= \delta/2$ as functions of $q$, showing 
the change from maximum (negative $F^{\prime \prime}$) to 
minimum (positive $F^{\prime \prime}$). 

\begin{figure}
\begin{center}
\includegraphics[width=6cm]{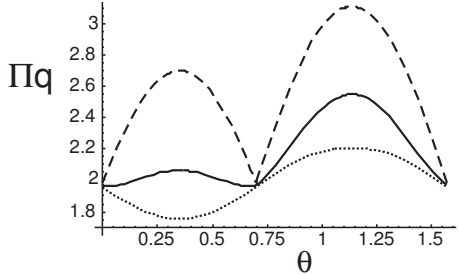}
\end{center}
\caption{Plot of $\Pi_q = R_q (A) R_q (B)$ as a function of 
$\theta$ for $\delta = 0.7$ and $q=0.5$ (dashed line), $q=1$ 
(solid line) and $q=1.5$ (dotted line).}
\end{figure}

\begin{figure}
\begin{center}
\includegraphics[width=6cm]{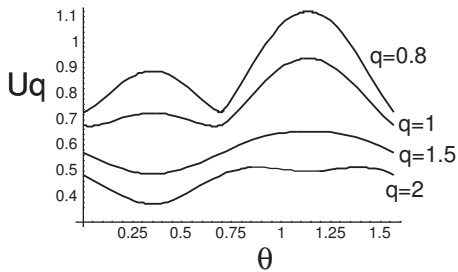}
\end{center}
\caption{Plot of $U_q = S_q (A) + S_q (B) + (1-q) S_q (A) S_q (B)$ 
as a function of $\theta$ for $\delta = 0.7$ with $q=0.8,1,1.5,2$.}
\end{figure}

\begin{figure}
\begin{center}
\includegraphics[width=6cm]{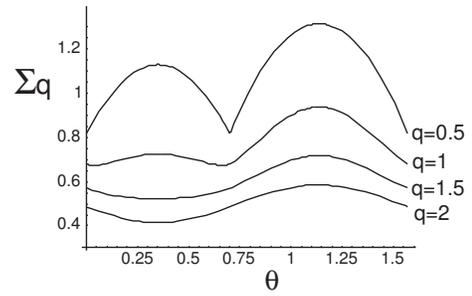}
\end{center}
\caption{Plots of $\Sigma_q = S_q (A) + S_q (B)$ as a function of 
$\theta$ for $\delta = 0.7$ and $q=0.5,1,1.5, 2$.}
\end{figure}

\begin{figure}
\begin{center}
\includegraphics[width=6cm]{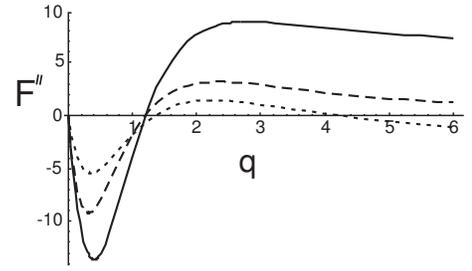}
\end{center}
\caption{Plot the second derivative at $\theta = \delta/2$ of 
$\Pi_q$ (solid line), $U_q$ (dashed line), and $\Sigma_q$ 
(dotted line) as functions of $q$ for $\delta = 0.7$.}
\end{figure}

Moreover, in Figs. 9 and 10 it can be appreciated that for 
$\Pi_q$ and $U_q$ the absolute maximum for low $q$ at $\theta 
= \delta/2 + \pi/4$ becomes a local minimum for larger $q$. 

\begin{figure}
\begin{center}
\includegraphics[width=6cm]{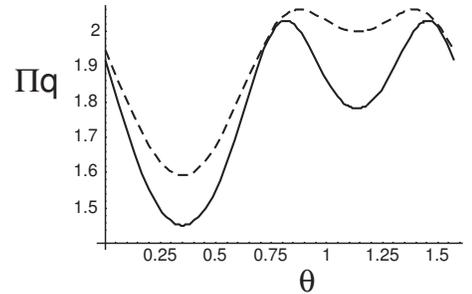}
\end{center}
\caption{Plot of $\Pi_q = R_q (A) R_q (B)$ as a function of 
$\theta$ for $\delta = 0.7$ and $q=2$ (dashed line) and $q=3$ 
(solid line).}
\end{figure}

\begin{figure}
\begin{center}
\includegraphics[width=6cm]{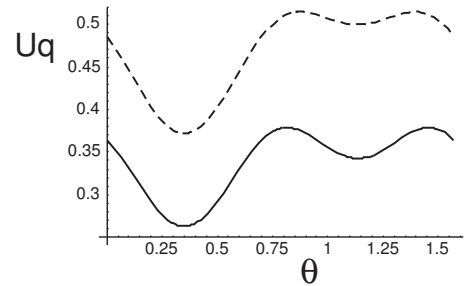}
\end{center}
\caption{Plot of $U_q = S_q (A) + S_q (B) + (1-q) S_q (A) 
S_q (B)$ as a function of $\theta$ for $\delta = 0.7$ and 
$q=2$ (dashed line) and $q=3$ (solid line).}
\end{figure}

\textit{Discussion.--}
The above plots show that maximum uncertainty states can become 
minimum uncertainty states and vice versa, depending on the measure 
of uncertainty employed, even with choices within the same family 
of measures. To some extent is natural that different measures lead 
to different extremes. However, it seems paradoxical and 
counterintuitive that the conclusions may be contradictory to the 
extent of exchanging maxima and minima.

Despite its long history, uncertainty relations may still provide 
surprising results worth investigating. For example, recently it 
has been put forward that there are fluctuations measures that 
seemingly lead to no uncertainty relation for complementary 
observables \cite{yR,ZZ}.

A. L. acknowledges support from Project No. FIS2008-01267 
of the Spanish Direcci\'{o}n General de Investigaci\'{o}n del 
Ministerio de Ciencia e Innovaci\'{o}n, and from Project 
QUITEMAD S2009-ESP-1594 of the Consejer\'{\i}a de Educaci\'{o}n
de la Comunidad de Madrid.


\begin{thebibliography}{00}

\bibitem{var}
J. Hilgevoord, Am. J. Phys. \textbf{70}, 983 (2002); 
G. N. Lawrence, Laser Focus World \textbf{30}, 109 (1994);
J. \v{R}eh\`{a}\v{c}ek and Z. Hradil, J. Mod. Opt. \textbf{51}, 
979 (2004).

\bibitem{ypra}
A. Luis, Phys. Rev. A \textbf{64}, 012103 (2001); 
\textbf{67}, 032108 (2003).

\bibitem{fase}
J. M. L\'{e}vy-Leblond, Ann. Phys. (N.Y.)
\textbf{101}, 319 (1976);
E. Breitenberger, Found. Phys. \textbf{15}, 
353 (1985);
J. B. M. Uffink, Phys. Lett. A \textbf{108}, 59 
(1985);
J. M. L\'{e}vy-Leblond, \textit{ibid.} 
\textbf{111}, 353 (1985);
S. M. Barnett and D. T. Pegg, J. Mod. Opt. \textbf{36},
7 (1989);
Z. Hradil, Phys. Rev. A \textbf{46}, R2217 (1992); 
Quantum Opt. \textbf{4}, 93 (1992); 
T. Opatrn\'{y}, J. Phys. A \textbf{27}, 7201 (1994);
V. Pe\v{r}inov\'{a}, A. Luk\v{s}, and J. Pe\v{r}ina,
\textit{Phase in Optics} (World Scientific, Singapore, 
1998).

\bibitem{atv}
I. I. Hirschman, Am. J. Math. \textbf{79}, 152 (1957);
I. Bialynicki-Birula and J. Mycielski, Commun. Math. Phys. \textbf{44}, 129 (1975);
D. Deutsch, Phys. Rev. Lett. \textbf{50}, 631 (1983);
M. Hossein Partovi, \textit{ibid.} \textbf{50}, 1883 (1983); 
K. Kraus, Phys. Rev. D \textbf{35}, 3070 (1987);
H. Maassen and J.B.M. Uffink, Phys. Rev. Lett. \textbf{60}, 1103 (1988);
J. Sanchez, Phys. Lett. A \textbf{173}, 233 (1993); 
\v{C}. Brukner and A. Zeilinger, \textit{ibid.} \textbf{83}, 3354 (1999);
Phys. Rev. A \textbf{63}, 022113 (2001);
V. Majern\'{\i}k and E. Majerníkov\'{a}, Rep. Math. Phys. \textbf{47}, 381 (2001);
M. J. W. Hall, Phys. Rev. A \textbf{64}, 052103 (2001);
S. Massar and P. Spindel, Phys. Rev. Lett. \textbf{100}, 190401 (2008); 
S. Wehner and A. Winter, New J. Phys. \textbf{12}, 025009 (2010); 
I. Bialynicki-Birula and L. Rudnicki arXiv:1001.4668v1;
I. Urizar-Lanz and G. T\'{o}th, Phys. Rev. A \textbf{81}, 052108 (2010); 
P. S\'{a}nchez-Moreno, A. R. Plastino, and J. S. Dehesa, J. Phys. A \textbf{44},
065301 (2011);
A. E. Rastegin, \textit{ibid.} \textbf{44}, 095303 (2011).  

\bibitem{Ts}
C. Tsallis, J. Stat. Phys. \textbf{52}, 479 (1988);
E.M.F. Curado and C. Tsallis. J. Phys. A \textbf{24}, L69 (1991);
A. K. Rajagopal, Phys. Lett. A \textbf{205}, 32 (1995).

\bibitem{PP}
M. Portesi and A. Plastino, Physica A \textbf{225}, 412 (1996). 

\bibitem{Ry}
A. Renyi, "On the measures of entropy and information," in
Proceedings of the 4th Berkeley Symposium on Mathematics
and Staticstical Probability (University of California Press,
1961), Vol. 1, pp. 547–561;
U Larsen, J. Phys. A \textbf{23}, 1041 (1990);
I. Bialynicki-Birula, Phys. Rev. A \textbf{74}, 052101 (2006).

\bibitem{yR}
A. Luis, Opt. Lett. 31, 3644 (2006);
Phys. Rev. A \textbf{75}, 052115 (2007).

\bibitem{ZZ}
S. Zozor, M. Portesi, and C. Vignat,  Physica A \textbf{387}, 4800 (2008).

\bibitem{yomlh}
A. Luis, Phys. Rev. A \textbf{80}, 034101 (2009).


\end{thebibliography}
\end{document}